\documentclass[a4paper]{article}

\usepackage{INTERSPEECH2022}
\usepackage{hyperref}
\hypersetup{
    colorlinks=true,
    linkcolor=blue,
    filecolor=magenta,      
    urlcolor=cyan,
    pdftitle={Overleaf Example},
    pdfpagemode=FullScreen,
    }

\title{Improving Self-Supervised Learning-based MOS Prediction Networks}
\name{Bálint Gyires-Tóth, Csaba Zainkó}

\address{
  Department of Telecommunications and Media Informatics\\ Faculty of Electrical Engineering and Informatics\\ Budapest University of Technology and Economics \\
  Műegyetem rkp. 3., H-1111 Budapest, Hungary }
\email{\{toth.b, zainko\}@tmit.bme.hu}

\begin{document}

\maketitle
\begin{abstract}

MOS (Mean Opinion Score) is a subjective method used for the evaluation of a system's quality. Telecommunications (for voice and video), and speech synthesis systems (for generated speech) are a few of the many applications of the method. While MOS tests are widely accepted, they are time-consuming and costly since human input is required. In addition, since the systems and subjects of the tests differ, the results are not really comparable. On the other hand, a large number of previous tests allow us to train machine learning models that are capable of predicting MOS value. By automatically predicting MOS values, both the aforementioned issues can be resolved. 

The present work introduces data-, training- and post-training specific improvements to a previous self-supervised learning-based MOS prediction model. We used a wav2vec 2.0 model pre-trained on LibriSpeech, extended with LSTM and non-linear dense layers. We introduced transfer learning, target data preprocessing a two- and three-phase  training method with different batch formulations, dropout accumulation (for larger batch sizes) and  quantization of the predictions.

The methods are evaluated using the shared synthetic speech dataset of the first Voice MOS challenge. 
\end{abstract}

% wav2vec, pre-train, transfer-learning, batch with same size, padding left/right, lr-schedule, regression/classification, 0,125 kerekítés, LSTM hyperopt, dropout, accumlation (big batch size), f0/Mel spectrum (nem segített)

\section{Introduction}
In the majority of generative deep learning models the objective function is not or only weakly correlated with the subjective perception. Thus, human interaction is necessary to evaluate the models. For such purpose the Mean Opinion Score (MOS) \cite{recommendation2006vocabulary} method can be used.
MOS is considered precise when more opinions are collected per sample. The results of a smaller number of test subjects provide a statistically imprecise approximation of human perception.\footnote{The MUSHRA method (MUltiple Stimuli with Hidden Reference and Anchor) can be used with small numbers of test subjects, \cite{series2014method}. MUSHRA introduces best- and worst-quality samples as anchors and each model's output is presented at the same time on the same sample, allowing paired t-tests or repeated measures analyses of variance to be conducted.} A pseudo-random sample order is often used in subjective tests, ensuring that all samples are scored about the same number of times, while the order does not influence the test subjects, preventing perception bias.
The MOS test is often used for evaluating speech generation models (for instance, in terms of naturalness, intelligibility, and quality). Several automatic approaches have been developed to approximate the MOS values. Approximation models have been more and more data-driven as machine learning models have emerged. 
In 2022, the first VoiceMOS Challenge was held with the goal of elaborating methods for automatic prediction of synthetic speech MOS values \cite{voicemos2022}. Our work for the challenge is presented in this paper, which is also a general approach to MOS prediction.\footnote{The corresponding code can be found at: https://github.com/BME-SmartLab/DeepMOS}

\section{Related work}
%asr: deepspeech, deepspeech2, wav2letter, wav2letter++, jasper, quarznet, citrinet, conformer
%tts: specotgram: tacotrton2, glowtts, fastspeec2, fastpitch, talknet
%tts: vocoder: waveglow, squeezeawve, uniglow, melglow, hifigan

Deep learning is the primary technique in speech modeling, nowadays. It has dominated both automatic speech perception (including automatic speech recognition (ASR), speaker diarization, voice activity detection, speech command recognition, etc.) and generation (spectrogram generation and vocoder, i.e. text-to-speech synthesis (TTS)). 
In ASR, DeepSpeech \cite{hannun2014deep}, wav2letter \cite{collobert2016wav2letter}, Citrinet \cite{majumdar2021citrinet} are among the many end-to-end solutions. While DeepSpeech utilizes recurrent neuronal networks (RNN), the latter two are based on convolutional neural networks (CNN). For TTS, there are also a large number of different approaches: Tacotron2 \cite{shen2018tacotron2} uses both RNN and CNN layers to generate spectrogram from text, WaveNet \cite{van2016wavenet} is a convolution only neural vocoder, and HiFiGAN \cite{kong2020hifi} utilizes the generator of a Generative Adversarial Network for vocoding, where both the generator and discriminator are CNNs -- just to name a few. 
wav2vec fully convolutional models are trained with a contrastive, self-supervised manner for speech recognition by learning speech representations from raw audio on a pretext task \cite{schneider2019wav2vec} and fine-tuned to different downstream tasks later. The second version, wav2Vec 2.0 \cite{baevski2020wav2vec} masks latent representations of the raw waveform and solves a contrastive task over quantized speech representations. It is able to outperform semi-supervised methods. 
The use of deep learning has also become increasingly prevalent in synthetic speech quality estimation. 
\cite{yoshimura2016hierarchical} uses CNNs to predict the naturalness MOS values on utterance and on system levels. 
\cite{MOSlo2019mosnet} uses BLSTM, CNNs CNN-BLSTM for the quality assessment of voice conversion (VC), and also show the generalization capability of their approach. 
%The non-intrusive Quality-Net applies bidirectional Long Short-Term Memory (BLSTM)  \cite{fu2018quality}, trained on a dataset augmented with different kinds and amounts of noise. The results are highly correlated with PESQ \cite{rix2001pesq}. 
The well-speared latent representation clusters of wav2vec 2.0 for different speech samples are investigated with t-SNE, and the wav2vec 2.0 is extended with a 'judge' identifier for better predictions in \cite{MOStseng2021utilizing}.
Additionally, MBNet  utilizes the judge identities in the training dataset to model the subjects' bias \cite{MOSleng2021mbnet} and shows further improvements.
wav2vec 2.0 is also used for feature extraction by SSL-MOS \cite{MOScooper2021generalization}, and for modeling, the extracted features are averaged over time, and a single linear layer is used as a regressor. This was one of the challenge's baselines, and this was the basis of our research. We have chosen the wav2vec 2.0 approach, as there are numerous published pre-trained models (uni- and multi-language) on hundreds of hours of recordings. However, it is important to note that there are many outstanding approaches to speech processing, so other neural architectures and training methods used in ASR, TTS, or any additional neural speech modeling tasks might also be outstanding candidates.

%The challenge provided the following open-source MOS predictors as baselines: MOSA-Net \cite{MOSzezario2021deep}, and LDNet \cite{MOShuang2021ldnet}, which correspond to system IDs B01, B02 and B03, respectively. 
%SSL-MOS adds a simple linear finetuning layer for the MOS prediction task onto Fairseq self supervised learning (SSL)-based models for speech. 
%MOSA-Net uses cross-domain features such as spectral information, complex features, raw waveform data, and features extracted from SSL-based models to estimate multiple speech assessment metrics simultaneously. Originally developed towards noisy speech quality assessment, MOSA-Net serves as a good example to demonstrate how adapting such a system to this task is possible 
%LDNet does not rely on any external datasets, but its specialized model structure and inference method allow for utilizing individual ratings to facilitate listener-dependent modeling.

\section{Data exploration}

The details of the dataset used in this research is introduced in the VoiceMOS Challenge paper \cite{voicemos2022}. In the main track of the challenge English audio samples of 187 different TTS and VC systems are involved, 8 MOS values for each samples, 4,974; 1,066 and 1,066 samples in the training, validation and test sets, respectively. In the out-of-domain (OOD) track 136 labelled and 540 unlabelled Chinese TTS samples were used as training, and 136 as development, and 540 as test set. 

%It is critical, that the dataset is collected with an adequate methodology. If the approximations of the true subjective values are imprecise, the machine learning model would be unable to learn general context, but will over- or underfit. 

\begin{figure}[!b]
  \centering
  \includegraphics[width=0.8\linewidth]{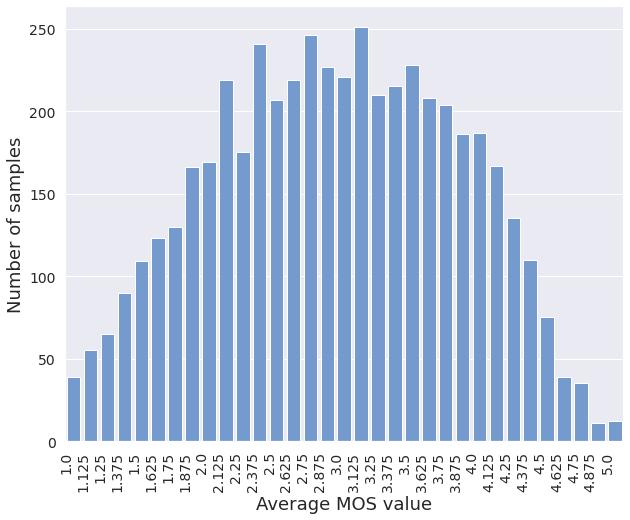}
  \caption{Distribution of the averaged MOS values.}
  \label{fig:MOS_mean_main}
\end{figure}

Prior to modeling, we believed it was important to examine the statistical characteristics of the dataset. We explored the dataset of the main track in this regard. Initially, we examined the distribution of the average MOS values. The result is depicted on Figure \ref{fig:MOS_mean_main}. The resolution of the MOS values is \( \frac{1}{8} = 0.125 \), as 8 MOS measurements are averaged for each samples in the main track. For OOD, the number of MOS values for one sample is 10..17, so the resolution in that case is varying between \( \frac{1}{10} \)...\( \frac{1}{17} \). The data and the figure show, that cca. $76\%$ of the MOS values lies in the $[2,4.125]$ range. The data of the $[1,2)$ and $(4.125,5]$ regions are about $16\%$ and $8\%$
of the complete dataset, respectively. As a result, in the main track's training dataset, the lower and higher MOS values are likely to be underrepresented. 
\label{section:data}

\begin{figure}[!t]
  \centering
  \includegraphics[width=0.95\linewidth]{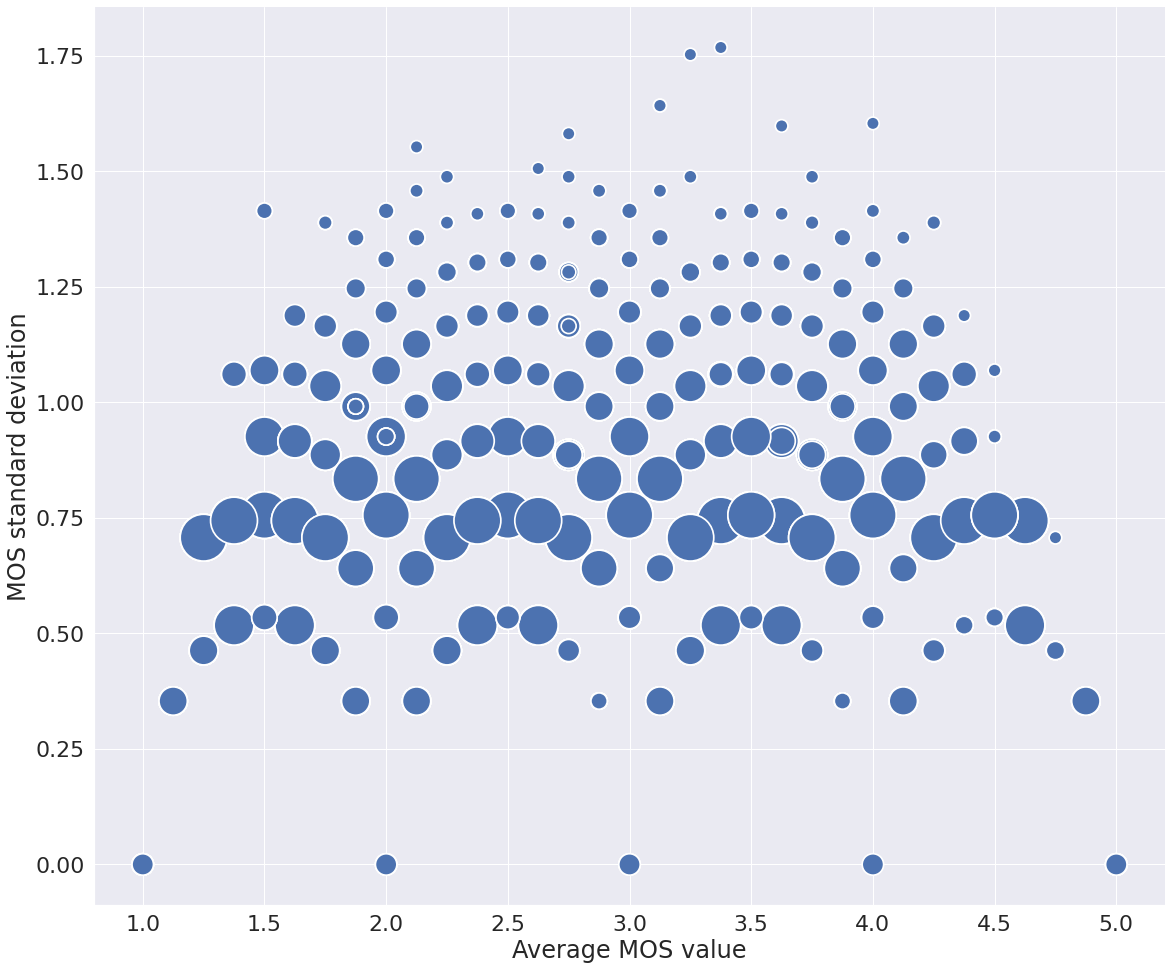}
  \caption{Correlation of the averaged MOS values and the MOS standard deviation, synthetic sample-wise. The size of the circles represent the amount of related data. }
  \label{fig:MOS_mean_std_main}
\end{figure}

Using the sample-wise MOS values, we next calculated the standard deviation and averages and illustrated the correlations, which is shown on Figure \ref{fig:MOS_mean_std_main}. Since the results of 8 samples were averaged, different people's subjective perceptions contributed to different values being voted for in some samples. The arcs show the degree of divergence between the perceptions of the participants. Lower arcs indicate consecutive MOS values (e.g. votes: 3,3,3,34,4,4,4, standard deviation: 0.5), while higher arcs demonstrate greater divergence in the consensus (e.g. votes: 1,1,2,3,3,4,5,5, standard deviation: 1.5). In addition, in case of round MOS values, a different level of standard deviation can also be examined. Consequently, a round average MOS value could result in either a perfect or devastating consensus. For instance averaged MOS value 3 can come from votes 3,3,3,3,3,3,3,3 (with 0 standard deviation), or 1,1,1,3,4,5,5,5 (with cca. 1.75 standard deviation). According to our analysis, standard deviation increases slightly for higher MOS values.

As a result of these observations, the variability of agreement and disagreement within MOS values generally establishes an irreducible error boundary for the modeling.

\section{Proposed methods}

Several steps were taken to improve the SSL-MOS approach. As the feature learner/extractor part a wav2vec 2.0 model pre-trained on the full 960 hours Librispeech corpus \cite{panayotov2015librispeech} was utilized. LSTM and fully connected non-linear layers followed the pre-trained wav2vec 2.0 submodel. 
Without further modification, training an architecture of this type seemed very unstable. The training is sensitive to the optimizer and its parameters, the batch compilations, the use of pre-trained models or the random seed. In the following section, we will discuss the details of the proposed models, and the training methodology. 

\subsection{Model architectures}

We approached the modeling task both with regression and classification models. In case of regression, the targets are the MOS value, while for classification, due to the 0.125 resolution (see Section \ref{section:data}) the samples are classified into one of the 33 categories (category 1: MOS=1, category 2: MOS=1.125, ..., category 33: MOS=5). The latter was inspired by \cite{van2016wavenet}, where an autoregressive waveform prediction was successfully realized as a 256-class classification approach.
The structure of the models is as follows (layers marked with \textbf{*} are introduced for regression only):

\begin{itemize}
    \item wav2vec 2.0 submodel: based on wav2vec 2.0 model pre-trained on the full Librispeech. 1024 dimensional output vectors are used.
    \item Dropout layer \textbf{*}: Dropout with probability $37.5\%$.
    \item LSTM layers: Two unidirectional LSTM layers are used with size 128. The hidden states were initialized with zero vector. The last element of the output is used only.
    \item Dropout layer \textbf{*}: Dropout with probability $75\%$.
    \item Dense layer: The size of layer is same as the number of LSTM cells, namely 128. 
    \item Activation: Sigmoid-Weighted Linear Units \cite{elfwing2018sigmoid}.
    \item Dropout layer \textbf{*}: Dropout with probability $75\%$.
    \item Dense layer: size of the layer corresponds to the number of outputs (1 for regression, 33 for classification. \item Softmax activation*

\end{itemize}

During inference, we combined both models and took the average of the predictions.

%The final model architecture is shown in the figure.
%
%TODO: ábra   Nem fog beférni, inkább kihagyom
%

\subsection{Training methodology}

\subsubsection{Batch compilation}
\label{section:batch}
Generally, the samples within a mini-batch are randomly selected during training. With this basic approach, large padding is required for shorter audio samples because of the substantially different sample sizes. It slows down the training process and increases memory usage for the same batch size. We resolved this limitation by sorting the training data by size, and forming mini-batches from successive samples. As a result, the length of the training data within a mini-batch barely differed, and little padding was needed. 

\subsubsection{Training the model}

In order to avoid trivial prediction (i.e. the input-independent mean value of the target training data) and to achieve a good convergence, several techniques had to be combined. 

We applied the batch compilation as described in the previous subsection with larger batch sizes. Since GPU memory only allowed smaller batch sizes (i.e. 8 for a 32GB GPU RAM), we introduced gradient accumulation with weight updates in every 10 steps. This is equivalent to a batch size of $10*8=80$, considering the gradient value.

We ran experiments with several optimizers, including Adam, AdamW, SGD, SGD with momentum and learning rate scheduling. Surprisingly, adaptive optimizers were not successful for this setup. SGD + momentum with or without learning rate scheduling gave the most stable convergence. As learning rate scheduler a cyclical learning rate scheduler with triangular form was utilized. The basic learning rate was 0.0005 and the maximum learning rate was 0.005. The length of a cycle was 200 iterations.

In the regression case, rather high dropouts were performing well (e.g. 75\% and above). 

The \textbf{regression model} was trained in one or two phases. After early stopping was called, we inspected the predictions. If the predictions were not covering to complete 1...5 MOS range (e.g. values below 1.5 and above 4.5 were missing), then we continued the training with unchanged settings over and over again. 

The predictions of the model was quantized to the 0.125 resolution in the main track (see Section \ref{section:data}).

%The \textbf{regression model} was trained in two phases. After the first phase of training (55 epoch), the prediction was  not uniform around the two endpoints, with no prediction of a value above 1 and no prediction above 4.5. In the second phase, the best model of the first phase was further trained with unchanged settings. Our experience is that the success of training is highly dependent on random initialization (random seed), so we selected the final model from several runs with the same parameters. It run only 2 epochs. 

Our aim with the \textbf{classification model} was to handle the 'missed' regions, that had higher errors with the regression model (i.e. near to MOS values 1.0 and 5.0). In our experiments, it was more difficult to train the classifier than the regression model. Similar learning methods used for the regression model did not converge. Therefor, we introduced transfer learning to the regression model with a three-phase training method.

Weights of the best regression model were used, except the final layer. The batch size was 100, SGD with momentum was used as optimizer and for loss function categorical cross entropy was utilized. The error was weighted by the reciprocal of the occurrence of the categories (Fig. \ref{fig:MOS_mean_main}) in order to handle class imbalances. 

In first phase, learning rate was set to 0.0005 and the wav2vec 2.0 submodel part (which was trained further in the regression model from the pre-trained version) was frozen. 
In the second phase the batch size was increased by $1.5x$ and a learning rate was decreased to $20\%$ of the original value. 
In the third phase, all layers of the model were unfrozen and taught and the batch size was decreased to 8.

%\subsubsection{Filtered training data}
%Skijump scoring
%
%EZ lehet, hogy kimarad.

\subsection{Out-of-domain model}

OOD model is a fine-tuned version of the regression model trained on the main track's data.  The model was trained with a learning rate of 0.0001 and a batch size of 10. % for 223 epochs.

\section{Evaluation and results}
% The given model is a hyperparameter optimized model tested with different LSTM cell sizes, number of LSTM layers, uni- and bidirectional LSTMs and dropout values. Many trainings failed, in many cases the model only predicted the average of the training set. The classifier was even more sensitive to the train parameters, the dropout layers had to be omitted compared to the previous model.
%  When evaluating the models, we measured that each model performs worse than the average of the two model outputs. The prediction of the final model is the average of the Regression and Classification models' prediction.

% Dropout layers were used after the LSTM layer and between the three layers to avoid over-fitting and to generalize. The best results were achieved at 0.375, 0.75, 0.75.

% Our experience is that the success of training is highly dependent on random initialization (random seed), so we selected the final model from several runs with the same parameters. It run only 2 epochs. 

% The first phase ran up to 396 epochs and then all layers except the SSL model were allowed to learn. This second phase ran for 2 epochs with a batch size of 150 and a learning rate of 0.0001.

% In the third phase, all layers were taught. With a batch size of 8 and a learning rate of 0.0001, the best result was achieved after 174 epochs. 

\begin{figure}[t]
  \centering
  \includegraphics[width=0.95\linewidth]{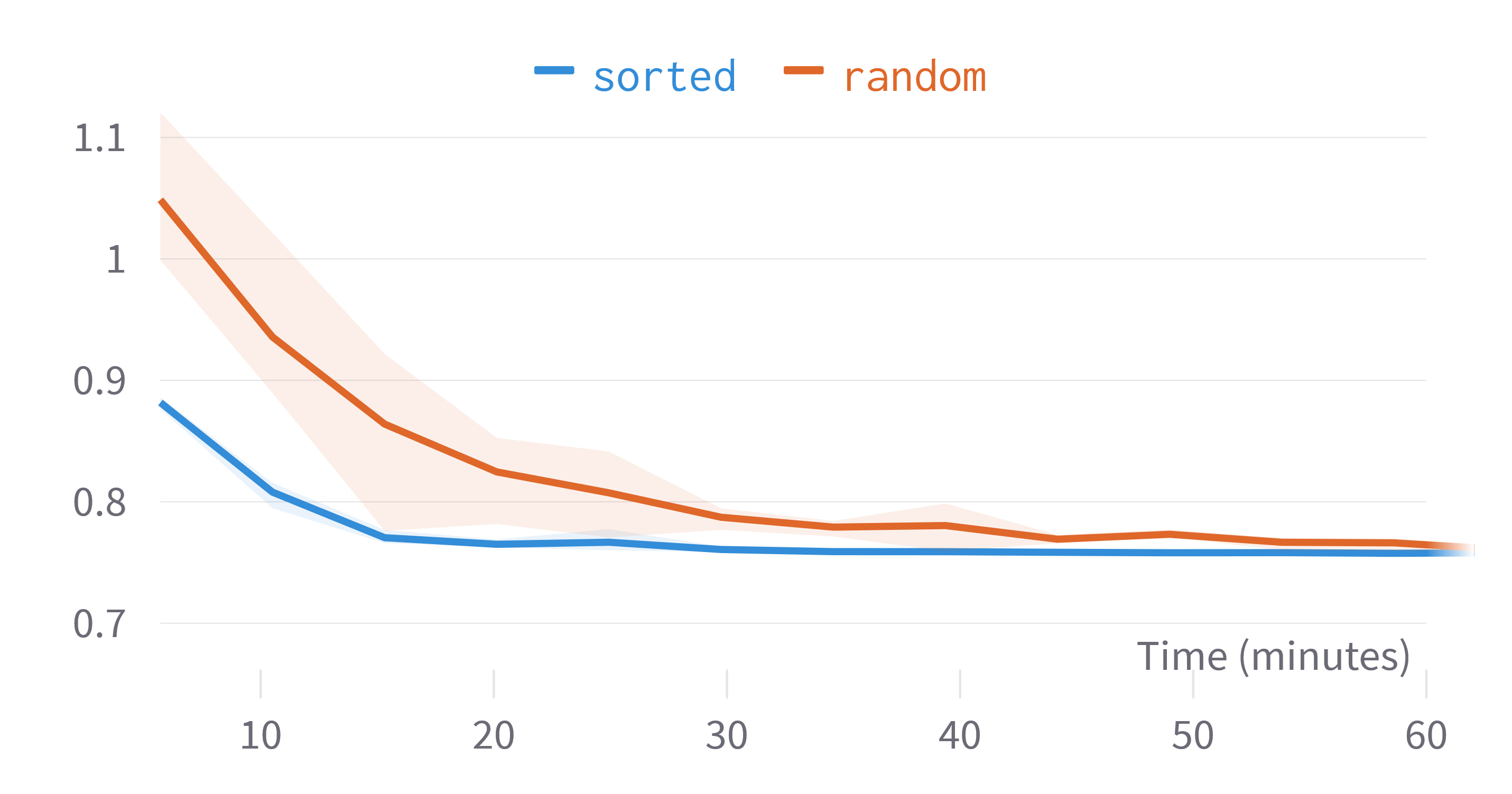}
  \caption{Validation loss with different batch compilation strategies.}
  \label{fig:batch_val_loss}
\end{figure}

We tested first the batch compilation technique, introduced in Subsection \ref{section:batch}. 3-3 trainings were performed. In half of the cases the samples within a mini-batch were randomly selected (referred to as \textit{random}), while in the other half the introduced method was used (referred to as \textit{sorted}). The only difference between the trainings apart from the batch compilation was the random seed (thus, the weights' initial values). Figure  \ref{fig:batch_val_loss} shows the results loss. With the proposed batch compilation approach it took significantly less time for the models to converge. 

\begin{figure}[t]
  \centering
  \includegraphics[width=0.7\linewidth]{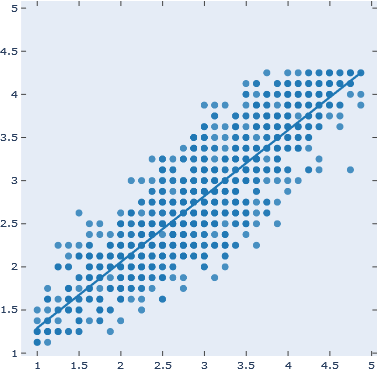}
  \caption{The correlation between prediction (vertical) and the ground truth (horizontal) of the regression model on the validation set of the main track.}
  \label{fig:regplot2}
\end{figure}

\begin{figure}[t]
  \centering
  \includegraphics[width=0.7\linewidth]{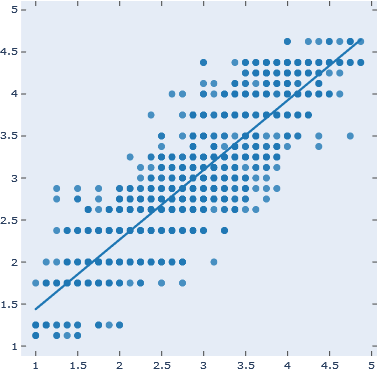}
  \caption{The correlation between prediction (vertical) and the ground truth (horizontal) of the classification model on the validation set of the main track. }
  \label{fig:regplot3}
\end{figure}

In the challenge, there were a training and an evaluation phase. For the former, both training and validation (also referred to as development set in the challenge) data was provided with ground truth values. In the evaluation phase, only the inputs were published, and on the online platform participants had a limited possibility to evaluate their predictions (maximum three uploads).
For selecting best model during the training phase, we used the validation set.
The evaluation was carried out in two levels: (TTS) system- and utterance-level. Four criteria were introduced in the challenge: mean squared error (MSE), Linear Correlation Coefficient (LCC), Spearman Rank Correlation Coefficient (SRCC), and Kendall Tau Rank Correlation (KTAU). 
The results on the validation set are shown in the Figure \ref{fig:regplot2} and \ref{fig:regplot3} for the regression and classification models, respectively. 

According to Figure \ref{fig:regplot2}, the regression model is predicting consequently lower values then the ground truth. The standard deviation of the error seems to be near to what was suspected in the data exploration part (subsection \ref{section:data}) as irreducible region. 

Inspecting Figure \ref{fig:regplot3} it can be concluded, that the classification model is not able to handle all classes properly. However, 'missed' regions of the regression model (MOS values near to 1.0 and 5.0) were handled better with the classifier. 

%\begin{figure}[t]
%  \centering
%  \includegraphics[width=0.95\linewidth]{figs/lstm_no_learn.png}
%  \caption{Validation loss of the biLSTM based model. TODO: rename lines}
%  \label{fig:lstm_no_learn}
%\end{figure}

%\begin{figure}[t]
%  \centering
%  \includegraphics[width=0.95\linewidth]{figs/firstphase.png}
%  \caption{Validation regplot}
%  \label{fig:regplot}
%\end{figure}

\subsection{Training and evaluation phases}

During the training phase, the models were evaluated primarily on the basis of the validation loss value. The best validation (L1) loss was 0.322 for the regression and 0.333 for the classification model. Combining the two models achieved better results. Consequently, the final predictions were combined. Table \ref{tab:valloss} shows the achieved system-level SRCC scores of the models on the validation set. 

\begin{table}[th]
  \caption{Training phase scores of the main track on the validation data.}
  \label{tab:valloss}
  \centering
  \begin{tabular}{l c} 
 \hline
 Model & SRCC\\ 
 \hline
 Regression  & 0.9468 \\ 
 %\hline
 Classification & 0.9440 \\
 %\hline
 Combined & 0.9498 \\
 %\hline
 Corrected & 0.9502 \\
 \hline
 
\end{tabular}
\end{table}

All models had difficulty handling values below 1.3 and above 4.2, so a correction factor was applied: -0.05 to the lower, and +0.25 to the higher region. The corrected model gave the best score, as shown in the table.

In evaluation phase the best model  obtained in the training phase was used. 

\subsubsection{Evaluation phase - main track}

Our model was able to predict the English samples in the Main track according to the values given in Table 
\ref{tab:main}. Our approach achieved better results than the baseline solution. 

\begin{table}[!t]
  \caption{Main track scores and achieved positions.}
  \label{tab:main}
  \centering
\begin{tabular}{l c c} 
 \hline
Score Name & Score Value & Position in the competition \\ 
 \hline
 System \hphantom{12}SRCC  & 0.9294 & \hphantom{0}9\\ 
 %\hline
 System \hphantom{12}MSE & 0.1298 & 10 \\
 %\hline
 Utterance SRCC & 0.8841 & \hphantom{0}8\\
 %\hline
 Utterance MSE & 0.2129 & 12\\
 \hline
 
\end{tabular}

\end{table}
\subsubsection{Evaluation phase - out-of-domain track}

For the OOD track, we achieved better positions in the competition. The overall result shows that the task was 'easier'. Since we focused on the main track, our OOD model is just a fine-tuned version of the English version. With this solution we achieved a result close to the baseline. Table \ref{tab:ood} shows the results.
%% Ha keves a hely, akkor a OOD es MAIN mehet egy tablazatba
\begin{table}[!t]
  \caption{OOD track scores and achieved positions.}
  \label{tab:ood}
  \centering\begin{tabular}{l c c} 
 \hline
Score Name & Score Value & Position in competition \\ 
 \hline
 System \hphantom{22}SRCC  & 0.9726 & 3\\ 
 %\hline
 System \hphantom{22}MSE & 0.0537 & 3 \\
 %\hline
 Utterance SRCC & 0.8746 & 3\\
 %\hline
 Utterance MSE & 0.1887 & 3\\
 \hline
 
\end{tabular}
\end{table}

\section{Summary}

In this paper we presented a solution for a MOS prediction competition. We used the training (English) dataset when developing the method, but throughout the teaching process we kept in mind only to use methods that were language-independent. The performance of our method is close to the best results of the competition and performed well on the Chinese TTS samples without substantial modification. Our plan is to test the method in other languages as well. 
Being near or within the irreducible error region of MOS values allow the reduction of the number of time- and resource-intensive human tests during the development of TTSs. It might also introduce a new era of perceived quality approximation-based TTS and VC model development. 

\section{Acknowledgements}
The research reported in this paper and carried out at BME has been supported by the National Laboratory of Artificial Intelligence funded by the NRDIO under the auspices of the Ministry for Innovation and Technology.
We thank for the usage of ELKH Cloud GPU infrastructure  (\url{https://science-cloud.hu/}) that significantly helped us achieving the results published in this paper. We gratefully acknowledge the support of NVIDIA Corporation with the donation of the NVIDIA GPU also used for this research.

\bibliographystyle{IEEEtran}

\bibliography{mybib}

% \begin{thebibliography}{9}
% \bibitem[1]{Davis80-COP}
%   S.\ B.\ Davis and P.\ Mermelstein,
%   ``Comparison of parametric representation for monosyllabic word recognition in continuously spoken sentences,''
%   \textit{IEEE Transactions on Acoustics, Speech and Signal Processing}, vol.~28, no.~4, pp.~357--366, 1980.
% \bibitem[2]{Rabiner89-ATO}
%   L.\ R.\ Rabiner,
%   ``A tutorial on hidden Markov models and selected applications in speech recognition,''
%   \textit{Proceedings of the IEEE}, vol.~77, no.~2, pp.~257-286, 1989.
% \bibitem[3]{Hastie09-TEO}
%   T.\ Hastie, R.\ Tibshirani, and J.\ Friedman,
%   \textit{The Elements of Statistical Learning -- Data Mining, Inference, and Prediction}.
%   New York: Springer, 2009.
% \bibitem[4]{YourName17-XXX}
%   F.\ Lastname1, F.\ Lastname2, and F.\ Lastname3,
%   ``Title of your INTERSPEECH 2022 publication,''
%   in \textit{Interspeech 2022 -- 23\textsuperscript{rd} Annual Conference of the International Speech Communication Association, September 18-22, Incheon, Korea, Proceedings, Proceedings}, 2022, pp.~100--104.
% \end{thebibliography}

\end{document}